\documentclass{article} 
\usepackage{hyperref,psfrag}
\usepackage{url}
\usepackage{url,cite}
\usepackage{hyperref}
\usepackage{times}
\usepackage[square,sort,comma,numbers]{natbib}
\usepackage{natbib}
\usepackage[outdir=./]{epstopdf}
\usepackage{graphicx,float,pgfplots,wrapfig,sidecap,lipsum}
\usepackage{tabularx}
\usepackage{booktabs}
\usepackage{paralist}
\usepackage{algorithm}
\usepackage[noend]{algorithmic}
\usepackage[pass]{geometry}
\usepackage{amsfonts,amsthm,amsmath,amssymb}
\usepackage{enumitem}
\usepackage{xcolor}
\usepackage[font=normal,labelfont=bf]{caption}
\usepackage{psfrag} 
\usepackage{siunitx}
\usepackage{subfig}

\usepackage{tikz}

\usetikzlibrary{fit}
\usetikzlibrary{calc,shapes}
\usetikzlibrary{decorations.pathmorphing} 
\usetikzlibrary{fit}					
\usetikzlibrary{backgrounds}	

\newcommand{\Dir}{\mathsf{Dir}}
\newcommand{\image}{I}
\newcommand{\filter}{F}
\newcommand{\map}{M}
\newcommand{\fro}{\mathsf{F}}
\newcommand{\tha}{^\mathsf{th}}

\newcommand{\bp}{\begin{psfrags}}
\newcommand{\ep}{\end{psfrags}}
\newcommand{\bc}{\begin{center}}
\newcommand{\ec}{\end{center}}

\usepackage{fullpage}
\usepackage{bbm}

\title{Discovering Neuronal Cell Types and Their Gene Expression Profiles Using a Spatial Point Process Mixture Model}

\author{
	Furong Huang \thanks{UC Irvine, furongh@uci.edu} 
	\and
	Animashree Anandkumar \thanks{UC Irvine, a.anandkumar@uci.edu}
	Christian Borgs \thanks{Microsoft Research, Christian.Borgs@microsoft.com} 
	\and
	Jennifer Chayes \thanks{Microsoft Research, jchayes@microsoft.com} 
	\and
	Ernest Fraenkel \thanks{MIT, fraenkel@mit.edu} 
	\and
	Michael Hawrylycz\thanks{Allen Institute, MikeH@alleninstitute.org}
	\and
	Ed Lein\thanks{Allen Institute, EdL@alleninstitute.org}
	\and
	Alessandro Ingrosso\thanks{Politecnico di Torino, alessandro.ingrosso@polito.it}
	\and
	Srinivas Turaga\thanks{HHMI Janelia Research Campus, turagas@janelia.hhmi.org}	
	}
\date{}
\author{
Furong Huang 
\\
{\scriptsize UC Irvine}\\
\texttt{{\scriptsize furongh@uci.edu}} \\
\and
Animashree Anandkumar \\
{\scriptsize UC Irvine} \\
\texttt{{\scriptsize a.anandkumar@uci.edu}} \\
\and
Christian Borgs \\
{\scriptsize Microsoft Research} \\
\texttt{{\scriptsize Christian.Borgs@microsoft.com}} \\
\and
Jennifer Chayes \\
{\scriptsize Microsoft Research} \\
\texttt{{\scriptsize jchayes@microsoft.com}} \\
\and
Ernest Fraenkel \\
{\scriptsize MIT} \\
\texttt{{\scriptsize fraenkel@mit.edu}} \\
\and
Michael Hawrylycz\\
{\scriptsize Allen Institute}\\
\texttt{{\scriptsize MikeH@alleninstitute.org}}\\
\and
Ed Lein\\
{\scriptsize Allen Institute}\\
\texttt{{\scriptsize EdL@alleninstitute.org}}\\
\and
Alessandro Ingrosso\\
{\scriptsize Politecnico di Torino}\\
\texttt{{\scriptsize alessandro.ingrosso@polito.it}}\\
\and
Srinivas Turaga\\
{\scriptsize HHMI Janelia Research Campus}\\
\texttt{{\scriptsize turagas@janelia.hhmi.org}}\\
}

\def\fighome{.}

\begin{document}
\maketitle

\begin{abstract}
Cataloging the neuronal cell types that comprise circuitry of individual brain regions is a major goal of modern neuroscience and the BRAIN initiative. Single-cell RNA sequencing can now be used to measure the gene expression profiles of individual neurons and to categorize neurons based on their gene expression profiles. While the single-cell techniques are extremely powerful and hold great promise, they are currently still labor intensive, have a high cost per cell, and, most importantly, do not provide information on spatial distribution of cell types in specific regions of the brain. We propose a complementary approach that uses computational methods to infer the cell types and their gene expression profiles through analysis of brain-wide single-cell resolution in situ hybridization (ISH) imagery contained in the Allen Brain Atlas (ABA). We measure the spatial distribution of neurons labeled in the ISH image for each gene and model it as a spatial point process mixture, whose mixture weights are given by the cell types which express that gene. By fitting a point process mixture model jointly to the ISH images, we infer both the spatial point process distribution for each cell type and their gene expression profile. We validate our predictions of cell type-specific gene expression profiles using single cell RNA sequencing data, recently published for the mouse somatosensory cortex. Jointly with the gene expression profiles, cell features such as cell size, orientation, intensity and local density level are inferred per cell type.
\end{abstract}
\section{Introduction}
	\subsection{Motivations and Goals}
	The human brain comprises about one hundred billion neurons and one trillion supporting glial cells. These cells are specialized into a surprising diversity of cell types. The retina alone boasts well over 50 cell types, and it is an active area of research to perform a census of the various neuronal cell types that comprise the central nervous system. Many criteria have been used to categorize neuronal cell types, from neuronal morphology and connectivity to their functional response properties. Neurons can also be categorized based on the proteins they make. Immunohistochemistry has been used with great success for many decades to differentiate excitatory neurons from inhibitory neurons by labeling for known proteins involved in the synthesis and regulation of glutamate and GABA, the primary excitatory and inhibitory neurotransmitters respectively.
	
	More recently, there has been an effort to systematically measure the complete transcriptome of single neurons. Single-cell RNA sequencing (RNA-Seq) is an extremely powerful technique that can quantitatively determine the expression level of every gene that is expressed in individual neurons. This so-called transcriptome or gene expression / transcription profile can then be used
to define cell types by clustering. A recent study produced the most comprehensive census of cell types to date in the mouse somatosensory cortex and hippocampus by performing single-cell RNA-Seq on over 3000 neurons~\cite{zeisel2015cell}. While this study is quite exciting, tyring to replicate it for all brain regions might well require the equivalent of a thousand such experiments.  Thus, it is likely that the unprecedented insights that RNA-Seq can provide will be slow to arrive. More importantly, single cell sequencing methods are not currently able to capture the precise three-dimensional location of the individual neurons.

\begin{figure}[!htb]
\begin{center}
\tikzset{
  mybackground/.style={execute at end picture={
        \begin{scope}[on background layer]
          \draw[black!15,fill=black!5,rounded corners=1ex] (current bounding box.south west)
                    rectangle (current bounding box.north east);
          \node[draw,fill=olive,ellipse,anchor=west,inner sep=0.1pt,minimum width=0.5ex] at (current bounding box.north
                   west){#1};
        \end{scope}
    }},
}
\tikzstyle{background}=[rectangle,
                                                fill=gray!10,
                                                inner sep=0.001cm,
                                                rounded corners=5mm]
\begin{tikzpicture}
[scale=.9, nodestyle/.style={fill = blue!10, shape = rectangle, rounded corners, minimum width = 1cm},]
\small
\matrix [column sep=0 cm,row sep=1.2cm] {
\node[nodestyle, align=center](a){\includegraphics[width = 1.5in,height = 1.5in]{\fighome/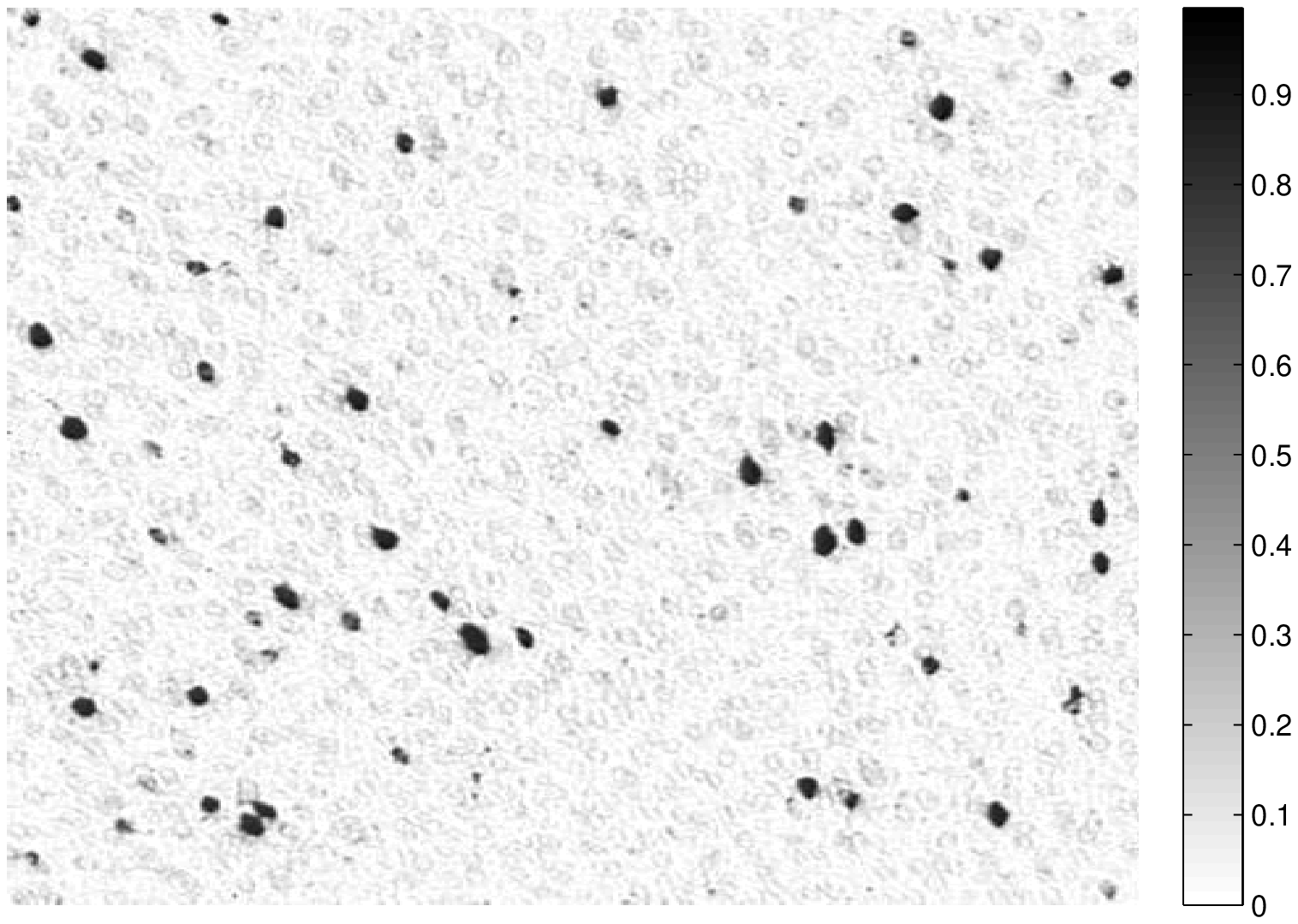}\\(a) Patch from gene Pvalb slice\\}; & \node[nodestyle, align=center](b){\bp \psfrag{orig}[Bc]{}\includegraphics[width = 1.5in, height =1.5in]{\fighome/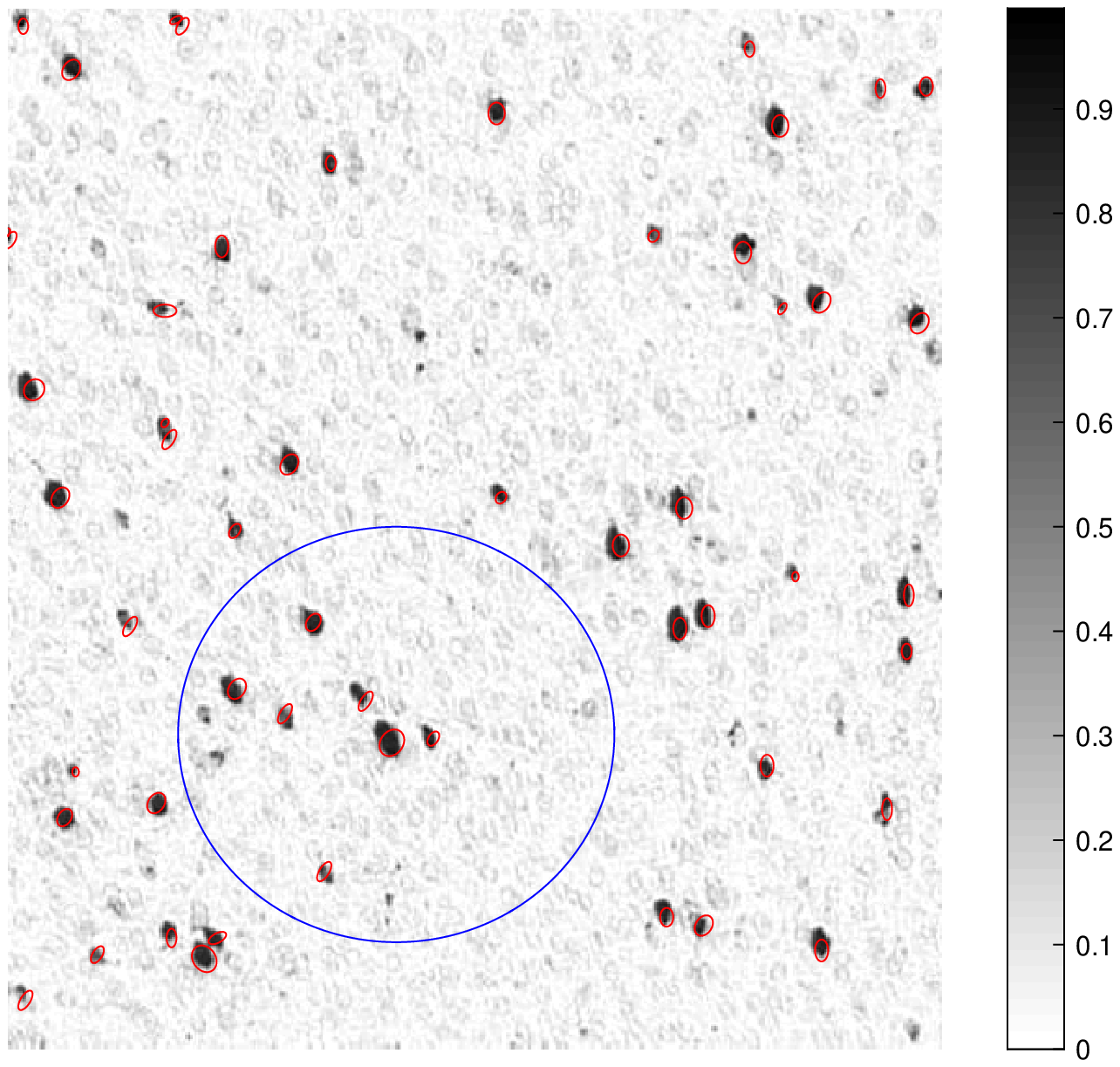} \ep \\(b) Cell detection and extraction of \\spatial point process features};\\

\node[nodestyle, align=center](c1){ 
{\psfrag{Number of Cells}[Bc]{\scriptsize{cell number}}
\psfrag{[diameter 1, diameter 2]}[Bc]{ }
\includegraphics[width=2.5in]{\fighome/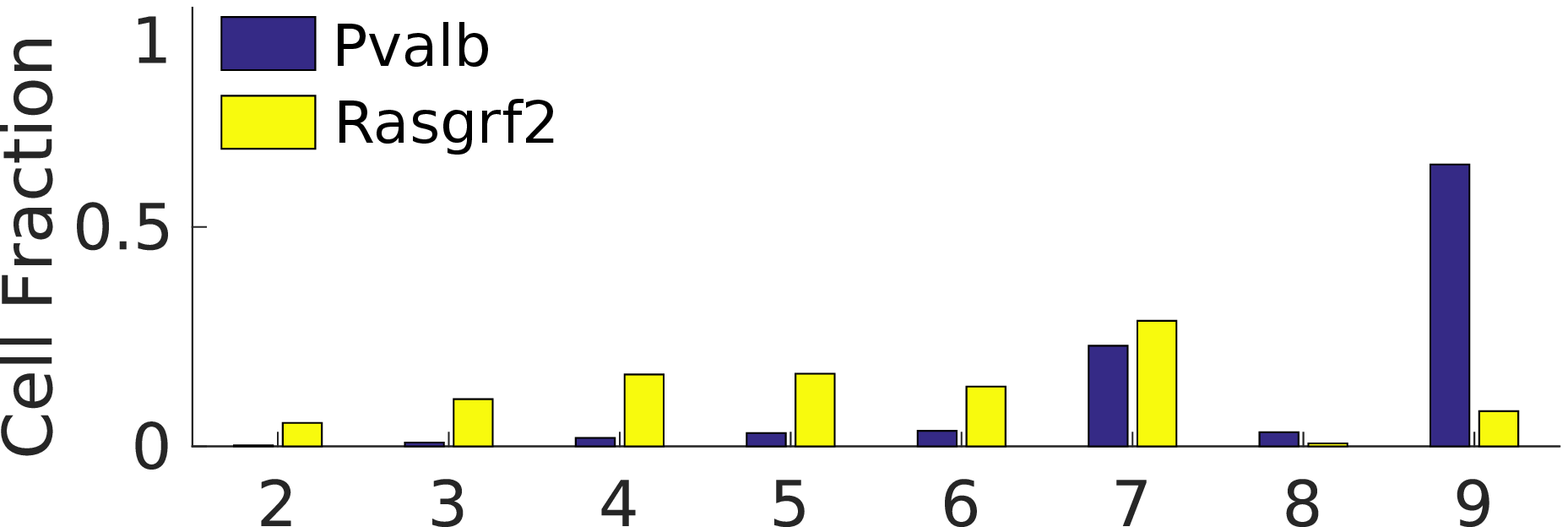}}\\
(c1) Size }; & 
\node[nodestyle, align=center](c2){
{ \psfrag{Number of Cells}[Bc]{\scriptsize{cell number}}
\psfrag{orientation}[Bc]{ }
\includegraphics[width=1.8in]{\fighome/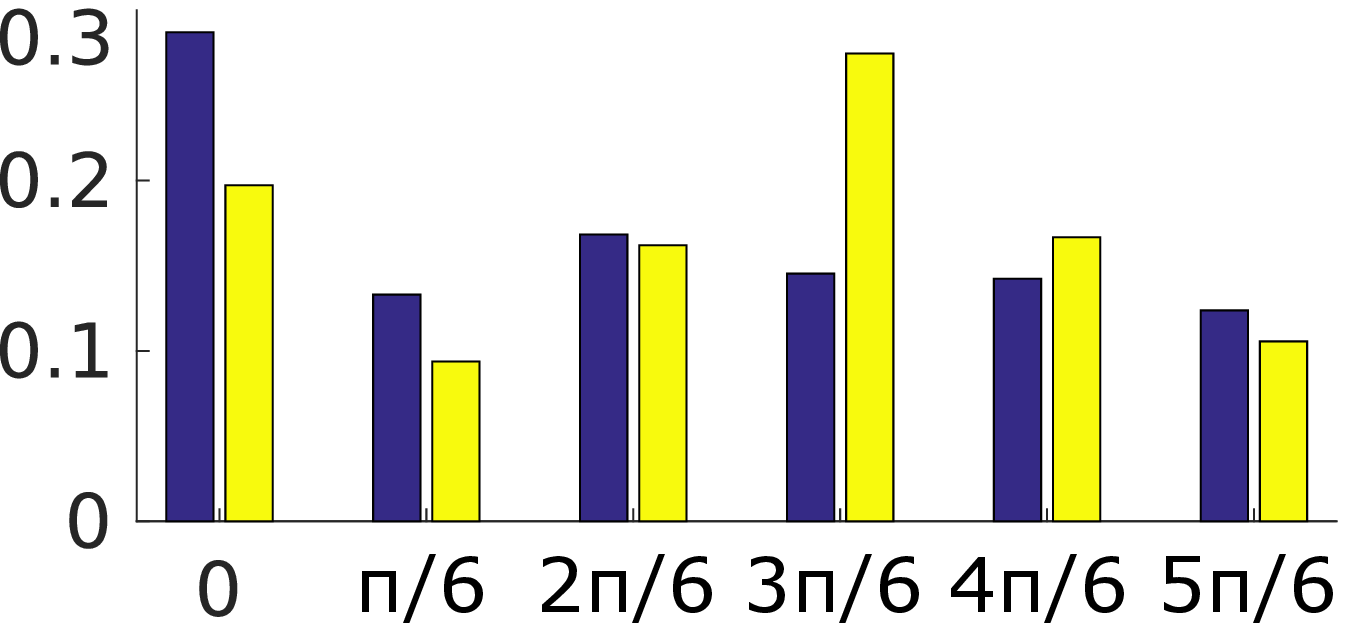}}\\
(c2) Orientation };\\

\node[nodestyle, align=center](c3){
{
\psfrag{Number of Cells}[Bc]{\scriptsize{cell number}}
\psfrag{Gene Profiles}[Bc]{}
\includegraphics[width=2.5in]{\fighome/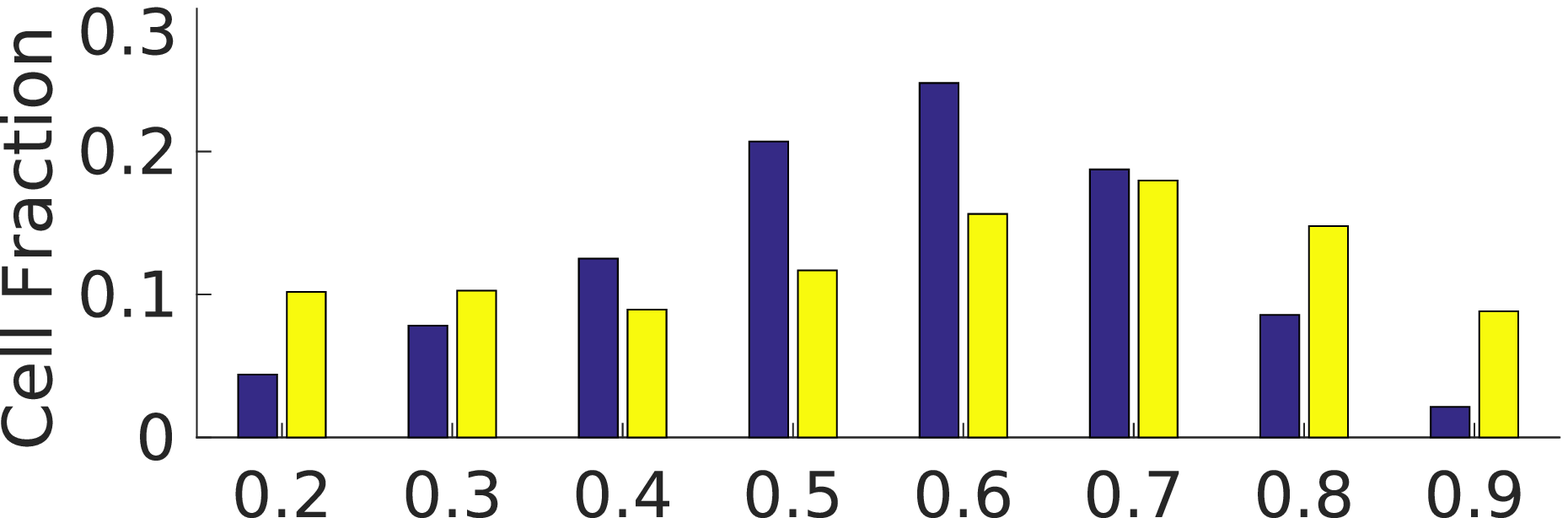}}\\
(c3)  Expression level }; &
\node[nodestyle, align=center](c4){
{
\psfrag{Number of Cells}[Bc]{\scriptsize{cell number}}
\psfrag{Cells within 100 um}[Bc]{}
\includegraphics[width=1.8in]{\fighome/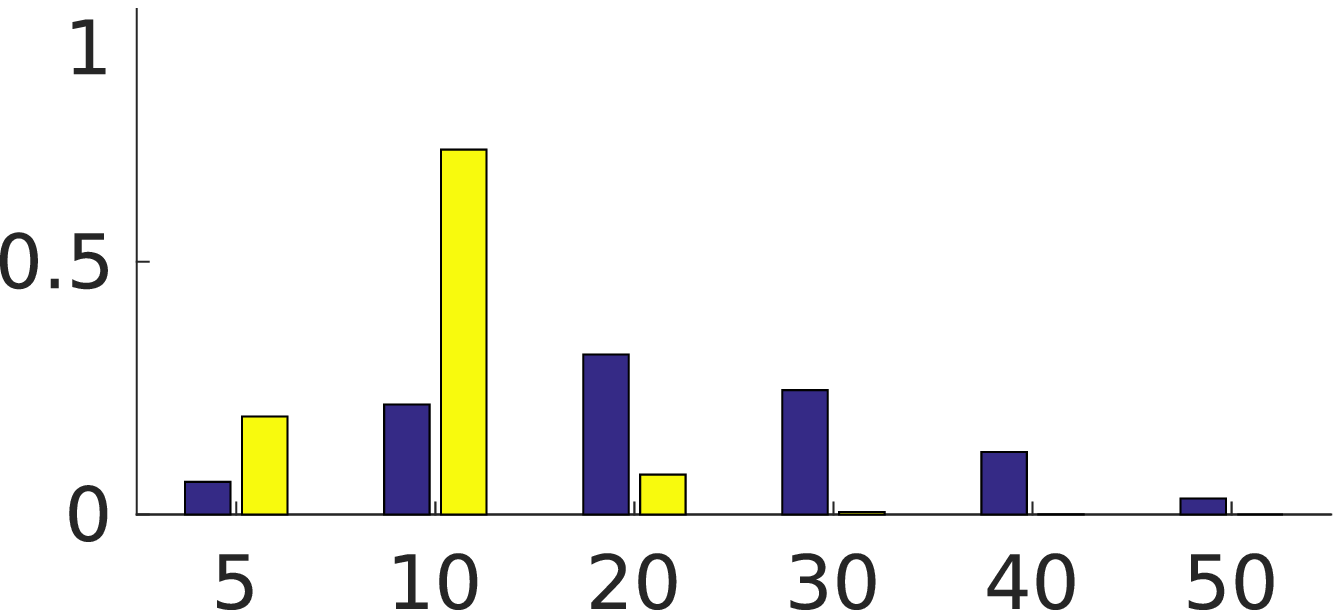}}\\
(c4)  Cell counts in \SI{100}{\micro\metre} radius}; \\

\node[nodestyle, align=center](d){(d) Point process histogram \\ representation: 
$[x^m_n]\in \mathbb{R_+}^{N_G \times N_F}$ };
& \node[nodestyle, align=center](f){(f) LDA model for inferring cell types}; \\ 
};
\begin{pgfonlayer}{background}
        \node [background,
                    fit=(a) (b),
                    label=above:\bf{\small Extract Point Process:}] (preprocess){};
        \node [background,
                    fit=(c1) (c2) (c3) (c4),
                    label=above:\bf{\small Joint Histogram:}] (jointHist){};
        \node [background,
                    fit=(d) (f),
                    label=above:\bf{\small Discover Cell Types:}] (mixturemodel){};
    \end{pgfonlayer}
    \path[->]
        (a) edge[thin,dotted] (b)
        (preprocess) edge[thin,dotted] (jointHist)
        (jointHist)   edge[thin,dotted] (mixturemodel)	
        (d) edge[thin,dotted] (f)
        ;
\end{tikzpicture}
\end{center}
\vspace{-1em}
\caption{\small Overview of the proposed framework - Discovering Neuronal cell Types via Un-mixing of  Spatial Point Process Mixtures.  (a) \& (b) An \emph{in situ} hybridization image for gene Pvalb along with detected cells. (c) Marginalized point process feature histograms for genes Pvalb and Rasgrf2. Note that size denotes the principal axis diameter. We have $N_G$ genes and 4d joint histogram with $N_F$ bins.}\label{fig:overview}
\vspace{-1em}
\end{figure}

	Here we propose a complementary approach that uses computational strategies to identify cell types and their spatial distribution by re-analysing data published by the Allen Institute for Brain Research.  The Allen Brain Atlas (ABA) contains cellular resolution brain-wide in-situ hybridization (ISH) images for 20,000 genes\footnote{ Although the Atlas contains ISH data for approximately 20,000 distinct mouse genes, we focus on the top 1743 reliable genes whose sagittal and coronal experiments are highly correlated.}. ISH is a histological technique that labels the mRNA in all cells expressing the corresponding gene in a manner roughly proportion to the gene expression level. An example of an ISH image can be seen in figure~\ref{fig:overview}(a).

	The ABA contains genome-wide and brain-wide ISH images of the adult mouse brain. These images were generated by slicing the brain into a series of \SI{25}{\micro\metre} thin sections and performing ISH. Image series of ISH performed for different genes come from different mouse brains, since ISH can only be performed for one gene at a time. The ISH image series for different genes were then computational aligned into a common reference brain coordinate system. Such data have been productively used to infer the average transcriptomes corresponding to different brain regions.

	It is commonly thought that the ABA cannot be used to infer the transcriptomes of individual cells in a given brain region since mouse brains cannot be aligned to the precision of a single cell. This is because there is individual variation in the precise number and location of neurons from brain to brain. However, we expect that the average number and spatial distribution of neurons from each cell type to be conserved from brain to brain, for a given brain area. More concretely, we might expect that parvalbumin-expressing (PV) inhibitory interneurons in layer 2/3 of the mouse somatosensory cortex comprise approximately 7\% of all neurons and have a conserved spatial and size distribution from brain to brain. We use this fact to derive a method for simultaneously inferring the cell types in a given brain region and their gene expression profiles from the ABA.
	
	We propose to model the spatial distribution of neurons in a brain as being generated by sampling from an unknown but consistent brain-region and cell-type dependent spatial point process distribution. And since each gene might only be expressed in a subset of cell types, an ISH image for a single gene can be thought of as a mixture of spatial point processes where the mixture weights represent the individual cell types expressing that gene. We infer cell types, their gene expression profiles and their spatial distribution by unmixing the spatial point processes corresponding to the ISH images for 1743 genes. This is in notable contrast to the information provided by single-cell RNA sequencing which can only measure the gene expression profile of individual cells to high accuracy but where, due to the destructive measurement process, all information about the spatial position and distribution of cell types is lost.
\subsection{Previous Work}
Allen Brain Atlas (ABA) \cite{lein2007genome} is a landmark study which mapped the gene expression of about 20,000 genes across the entire mouse brain. The ABA dataset consists of cellular high-resolution 2d imagery of \emph{in-situ} hybridized series of brain sections, digitally aligned to a common reference atlas. However, since the \emph{in-situ} images for each gene come from different mouse brains and since there is significant variability in the individual locations of labeled cells, it is not possible to register brain-wide gene expression at a resolution higher than about $250\mu m$. Therefore, the cellular resolution detail was down-sampled to construct a coarser 3d representation of the average gene expression level in $250\mu m\times 250\mu m\times 250\mu m$ voxels.

The coarse-resolution averaged gene expression representation has been widely used and analyzed to understand differences in gene expression at the level of brain region. Hawrylycz et al \cite{hawrylycz2011multi} analyzed the correlational structure of gene expression at this scale, across the entire mouse brain. However, due to the poor resolution of the average gene expression representation, it has proven challenging to use the ABA to discover the microstructure of gene expression within a brain region.  To address this issue from a complementary perspective, Grange et al~\cite{grange2014cell} used the gene expression profiles of 64 known cell-types, combined with linear unmixing to determine the spatial distribution of these known cell-types. However, such an approach can be confounded by the presence of cell-types whose expression profiles have yet to be characterized, and limited by the resolution of the averaged gene expression representation.

In contrast to previous approaches, we aim to solve the difficult problem of automatically discovering the gene expression profiles of cell-types within a brain region by analyzing the original cellular resolution ISH imagery. 
We propose to use the spatial distributions of labeled cells, and their shapes and sizes, which are a far richer representation than simply the average expression level in $250\mu m\times 250\mu m\times 250\mu m$ voxels. This spatial point process is then un-mixed to determine the gene expression profile of cell types.

Most previous work on unmixing point process mixtures adopted parametric generative models where the point process is limited to some distribution family such as Poisson or Gaussian~\cite{ji2009spatial,kottas2007bayesian}. However, since we are not interested in building a generative model of a point process, but rather care more about inferring the mixing proportions (gene expression profile), we take a simpler parameter-free approach. This approach models only the statistics of the point process, but is not a generative model, and so cannot be use to model individual points/cells.

\section{Modeling the Spatial Distribution of Cell-types Using Spatial Point Process Features}
Most analyses of the ABA \emph{in situ} hybridization dataset have utilized a simple measure of average expression level in relatively large $250\mu m \times 250\mu m \times 250\mu m$ voxels of brain tissue. Due to the large volume over which the expression level is averaged, such a representation cannot distinguish between  large numbers of cells expressing small amounts of RNA vs. small numbers of cells expressing large amounts of RNA. All information about the spatial organization of labeled cells, their shapes, sizes and spatial density are lost and summarized by a single scalar number. Here, we describe a more sophisticated representation of the labeled cells in an ISH image based on marked spatial point processes.
\subsection{The Marked Spatial Point Process Representation of ISH Images}
Our approach requires processing the high-resolution ISH images to detect individual labeled cells and their visual characteristics. We developed a cell detection algorithm described in the Supplementary section. Our algorithm additionally also estimates the expression level of each detected cell, its shape, size and orientation. Figure~\ref{fig:overview}(a) and Figure~\ref{fig:overview}(b) illustrate the results of our cell detection algorithm.

Since cell-types differ not only in terms of gene expression pattern, but also display a diversity of shapes, sizes and spatial densities, we sought to characterize these properties. We measured: (1) {\bf cell size} $s=[r_1,r_2]$: the radius in two principal directions of an ellipse fit to each cell; (2) {\bf cell orientation} $o$: the orientation of the first principle axis of the ellipse; (3) {\bf gene intensity level} $p$: intensity of labeling of a cell relative to the image background; (4) {\bf spatial distribution} $c$: the number of cells within a local area centered around the cell, which can be regarded as a measure of the local cell density.

The collection of detected cells within an atlas-defined brain region, along with their features, constitutes a marked spatial point process. This point process is considered ``marked'', because each point is characterized by the shape, size, expression level and local density features, in addition to just their location in space.
\subsection{A Model-free Approach to Representing Spatial Point Processes Using Joint Feature Histograms}
The statistical modeling of repulsive spatial point processes such as those that arise in biology is non-trivial, and many generative models such as determinantal point processes~\cite{kulesza2012determinantal}and Matern point processes have high computational complexity. But since we are not interested in directly modeling the individual labeled cells, but instead in modeling only their aggregate spatial statistics, and in inferring their gene expression profiles, we can take a simpler approach.

We use a \emph{joint histogram} simple statistics of the collection of detected cells to characterize the underlying point process from which they are drawn. This is an empirical moment approach which side-steps the need to carefully define a generative point process distribution.

As we describe in the next section, we propose to model the point process measured from the ISH image for each gene as a mixture of point processes belonging to individual cell-types. For this, we use a linear mixing model, the Latent Dirichlet Allocation model. The use of this model is greatly simplified if we carefully choose our feature representation such that the linear mixture of point processes results in a linear mixture of histogram statistics. This is clearly the case for the features we have chosen. For instance, if we sample equally from two point process distributions $P_1$ and $P_2$ with average densities of $d_1$ and $d_2$, the addition of these two point processes $P = P_1 + P_2$ results in the addition of the two densities $d = d_1 + d_2$. This is not the case for second order features, such as the distances to the nearest neighbors, which would have a more nonlinear relationship.

In figure~\ref{fig:overview}(c), we display marginal histograms corresponding to the joint histogram for two genes, Pvalb and Rasgrf2, which are well-known markers for a specific class of inhibitory and excitatory cortical neuronal cell-types respectively. 

\section{Un-mixing Spatial Point Processes to Discover Cell-types}
\subsection{Generative Model: A Variation of Latent Dirichlet Allocation }
The spatial point process histogram representation of the ABA ISH dataset results, for each brain region, is an $N_F\times N_G$ matrix $[x^m_n]$, where $N_F$ is the total number of histogram bins (henceforward called the number of histogram features) \footnote{Note that there are two types of \emph{features} -- the features characterizing each detected cell, and the features characterizing the collection of detected cells that constitute a single sample from a spatial point process}, $N_G$ is the number of genes, and $x^m_n$ is the number of cells expressing gene $n$ in histogram bin $m$.

We model the gene-spatial histogram matrix $[x^m_n]$ by assuming it is generated by a Variation of Latent Dirichlet Allocation (vLDA)~\cite{blei2003latent} model of cell types. This matrix factorization based latent variable model assumes that the ISH histograms are generated from a small number of cell-types, $K$, and each cell-type $i$ is associated with a type-dependent spatial point process histogram $h_i$ and a gene expression profile $\beta_i$.

Our generative model for each histogram bin $m$ (characterizing a particular bin in the size/ orientation/ gene profile/ spatial distribution) is as follows: Let $L^m=\sum_n^{N_G} x^m_n$ be the detected number of cells in the joint histogram bin $m$. For each cell $l$ in this bin, its cell-type $t$ is sampled from the multinomial distribution $h^m$. And given the cell-type $t$ of cell $l$, the genes $n$ expressed by this cell are sampled from a multinomial distribution given by the type-dependent gene expression profile/distribution $\beta^t$. For a given gene $n$ and histogram bin $m$, this generative process determines the number of cells that would be detected $x^m_n$.

We further place a Dirichlet prior over $h^m \sim \Dir(\alpha)$, with the concentration parameter $\alpha$ which determines the prior probability over the number of cell-types present in a given histogram bin $m$. This prior represents our prior knowledge of how many cell-types express each gene, and also how well our feature representation separates cells of different types into different histogram bins. In principle, we could generalize this to be a gene-specific prior, if we had such information available. We could also use $\alpha$ to incorporate information about our prior knowledge over the distribution of cells from each cell-type, for instance that excitatory neurons greatly outnumber inhibitory neurons in a roughly $5:1$ ratio. 

We now describe how we estimate the model parameters -- the cell-type specific multinomial gene expression profile $\beta$ and the cell-type specific spatial point process histogram $h$ from the gene-specific spatial point process histograms measured from the ISH images.
\subsection{Estimating the Cell-type Dependent Gene Expression Profile $\beta$}
After testing several estimation methods for the parameters of our model, we found that non-negative matrix factorization (NMF) performed well in estimating the cell-type specific gene expression profiles $\beta$, see Figure~\ref{fig:synthetic}. We solve the following optimization problem:
\begin{equation}
\min_{\beta,h} \quad \sum^{N_F}_m \sum^{N_G}_n (x^m_n - \sum^K_t h^m_t \beta^t_n L^m)^2, \quad
s.t. \quad \beta^t_n\ge 0 , \; \sum^{N_G}_n \beta^t_n=1, \; h^m_t \ge 0, \; \sum_t^K h^m_t=1
\end{equation}
Here, the non-negativity and sum-to-one constraints on $h^m_t$ and $\beta^t_n$ ensure that $h$ and $\beta$ result in properly normalized multinomial distributions. While this estimation procedure results in joint estimates for $h$ and $\beta$, it does not enforce the Dirichlet prior over $h$. So we refine our NMF-derived estimates for $h$ using variational inference \cite{blei2003latent}.
\subsection{Estimating the Cell-type Dependent Spatial Point Process Histogram $h$}
We use a standard maximum likelihood estimation procedure for $h$ \cite{blei2003latent}. Iteratively, we refine the  inference of the cell type membership $h^m \in \Delta_{k}$ under each joint histogram feature $m$. We update $h^{m}_i$ until convergence~\cite{smola2010architecture}.
\begin{equation}\label{eq:inferrence}
h^{m}_i \leftarrow \frac{1}{L^{m} + \sum_t^K\alpha_t} \sum\limits_{n=1}^{N_G} x^m_n \frac{h^{m}_i \beta^i_n}{\sum\limits_{l=1}^{K} h^{m}_l \beta^l_n} +\alpha_i , \ \forall i\in[K], m\in[N_F]
\end{equation}
Recall that the Dirichlet prior $\alpha$ encodes the number of cell-types that we expect on average to express each gene. We set $\alpha$ to be a symmetric Dirichlet with $\alpha_1=\alpha_2=\ldots=\alpha_K$, and $\sum_t \alpha_t=0.01$ for all cell-types $t$. In practice, we observe that our estimates of $h$ are fairly insensitive to the specific choice for $\alpha$ as long as $\sum_t \alpha_t$ is small enough. The smaller $\alpha$ is, the fewer cell-types expressing a given gene we expect to observe in a single histogram bin.

\section{Results and Evaluation}
\subsection{Implementation Details}
We tested our proposed cell-type discovery algorithm using the high-resolution \emph{in situ} hybridization image series for $1743$ of the most reliably imaged and annotated genes in the ABA. Individual cells were detected in the cellular resolution ISH images using custom algorithms (detailed in Supplementary Information). 
For each detected cell, we fit ellipses and extract several local features: (a) size and shape represented as the diameters along the principle axes of the ellipse, (b) orientation of the first principle axis, (c) gene intensity level as measured by the intensity of labeling of the cell body, and (d) the number of cells detected with-in a 100 $\mu m$ radius around the cell, which is a measure of the local cell density.
%
%
We aligned the ISH images to the ABA reference atlas and, for this paper, focused our attention on cells in the somatosensory cortex, since independent RNA-Seq data exist for this region the can be used to evaluate our approach. We computed joint histograms for the collection of cells found with-in the somatosensory cortex, resulting in a spatial point process feature vector of $N_F = 10010$ histogram bins per gene.
%
%

\begin{figure}[ht]
\subfloat[a][Validate NNMF Method]
{\begin{minipage}{0.49\textwidth}{
\begin{center}
\psfrag{Error Per Type}[Bl]{\tiny{Error Per Type}}
\psfrag{Number of Cell Types}[Bl]{\tiny{Number of Cell Types}}
\psfrag{Permuted beta error}[Bl]{\tiny{Permute $\beta$}}
\psfrag{Variational Inference beta estimation}[Bl]{\tiny{VI estimated $\beta$}}
\psfrag{NNMF beta estimation}[Bl]{\tiny{NNMF estimated $\beta$}}
\psfrag{NNMF beta robust estimation trueK 10}[Bl]{\tiny{NNMF robust}}
\includegraphics[width=\textwidth,height=1.4in]{\fighome/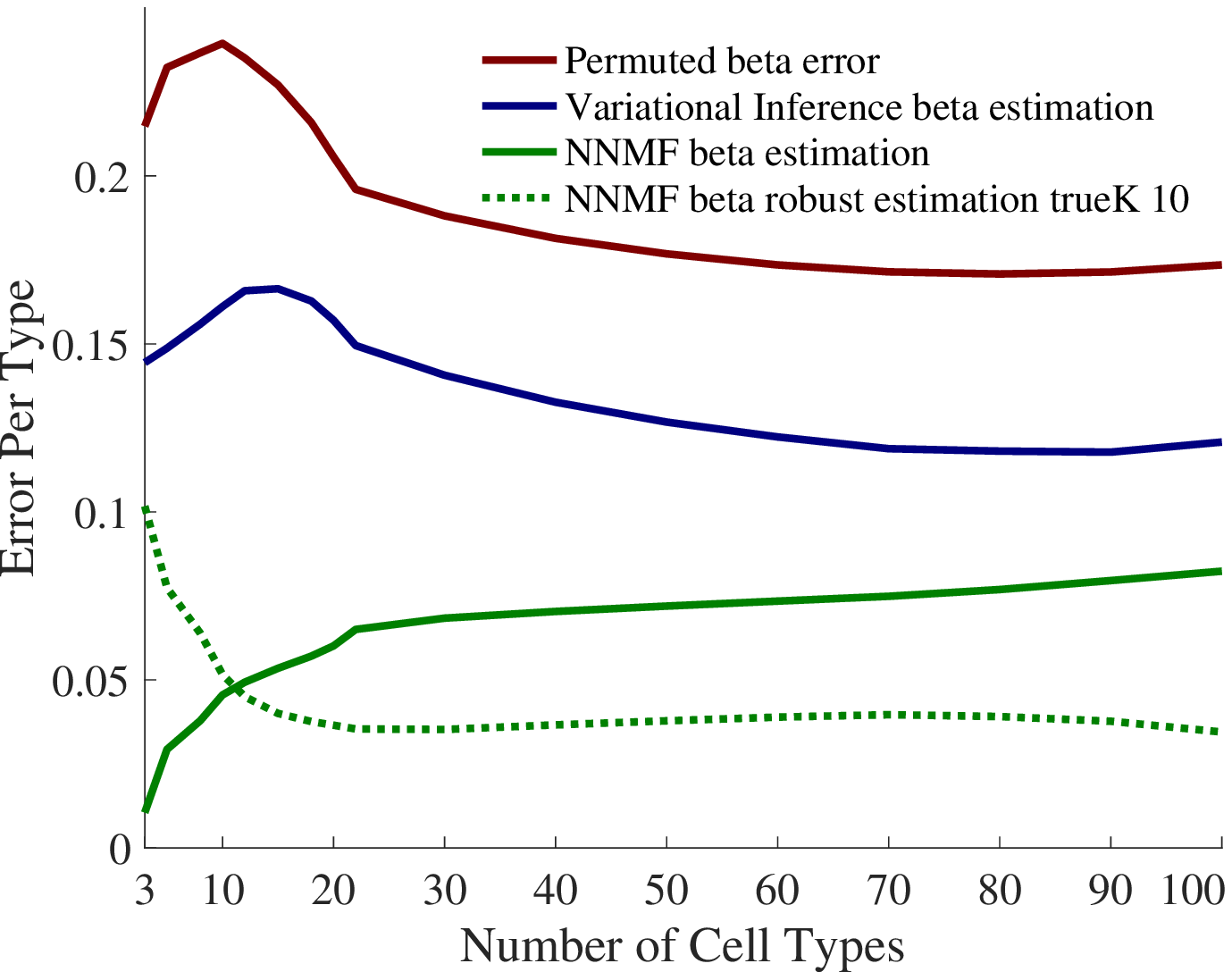}\label{fig:synthetic}
\end{center}
}
\end{minipage}}
\hfil
\subfloat[b][Validate Point Process Data]
{\begin{minipage}{0.49\textwidth}{
\begin{center}
\psfrag{Spatial Point Process Mixture ISH Data}[Bl]{\tiny{Spatial point process (ours)}}
\psfrag{Voxel Data}[Bl]{\tiny{Average expression level (baseline)}}
\psfrag{Spatial Point Process Mixture ISH Data (Permute)}[Bl]{\tiny{Spatial point process (ours, permuted)}}
\psfrag{Voxel Data (Permute)}[Bl]{\tiny{Average expression level (baseline, permuted)}}
\psfrag{Perplexity Score}[Bl]{\tiny{Perplexity Score}}
\psfrag{Number of Hidden Cell Types}[Bl]{\tiny{Number of Cell Types}}
\includegraphics[width=\textwidth]{\fighome/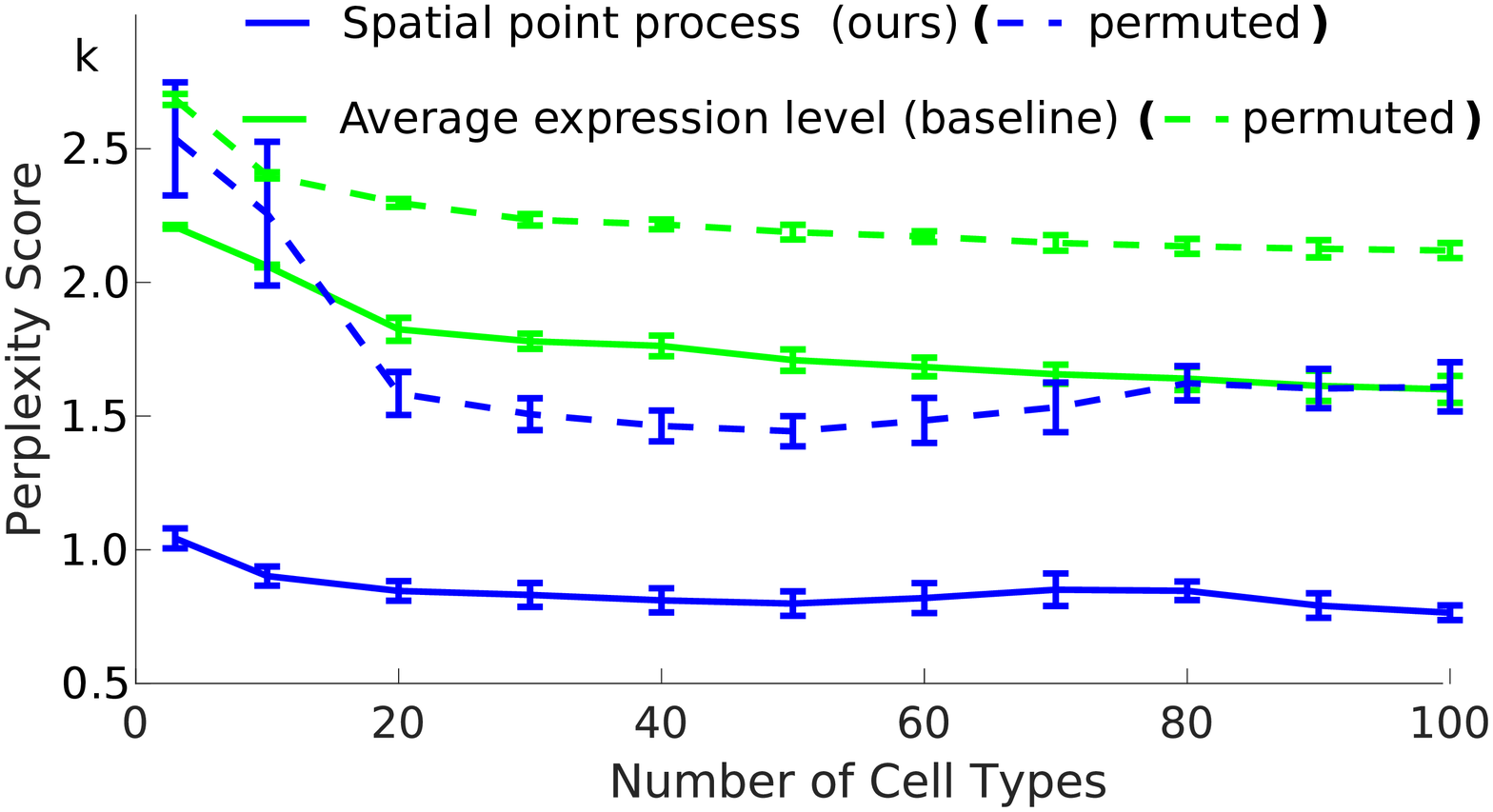}\label{fig:science}
\end{center}}
\end{minipage}}
\vspace{-0.5em}
\caption{(a) {Synthetic Experiment : comparison of  Non-negative Matrix Factorization (NNMF) with Variational Inference (VI) on simulated point process cell data using known gene expression profile $\beta$.  An additional robustness test of NNMF is done to see how good the algorithm is when a wrong number of cell types $K$ is input. A permutation test (shuffling the gene expression levels between cell) is done to access statistical significance. Comparing with permute test shows that our cell-types are significantly different from chance. Error per type is computed by pairing the columns of estimated $\widehat{\beta}$ with the columns of the ground-truth $\beta$.  }
(b) {Comparison of gene expression profiles recovered for cell-types in the somatosensory cortex by fitting an LDA model using spatial point process features (ours) vs the standard average gene expression level feature (baseline). Our features provide a significantly better match, with lower perplexity, to ground truth single-cell RNA sequencing derived transcriptomes.  A permutation test is done to access statistical significance. Perplexity is computed by matching to surrogate single-cell RNA transcriptomes by shuffling the gene expression levels between cells. Comparing with permute test shows that our cell-types are significantly different from chance.  }
}
\end{figure}

\paragraph{Synthetic experiment: } The vLDA model we proposed is then fit to $N_G \times N_F$ gene point process histogram matrix to estimate the cell-type gene expression profile matrix $\beta$ using the non-negative matrix factorization  (NNMF) algorithm.  The reason why we choose NNMF over Variational Inference (which is a popular approach for LDA) for $\beta$ estimation is that NNMF produces more accurate $\beta$ estimation in simulated data, illustrated in Fig~\ref{fig:synthetic}. In the synthetic experiment, we simulate point process data ( with some predefined golden standard $\beta$) and use the data to  estimate $\widehat{\beta}$. The errors were computed after pairing the estimated columns of $\beta$ with a closest golden standard $\beta$ column via hypothesis testing. Note that the columns of $\beta$ are normalized to 1, so the errors are bounded. 
\vspace{-2em}

\subsection{Evaluating Cell-type Gene Expression Profile Predictions}
A recent study performed single-cell RNA sequencing on $1691$ neurons isolated from mouse somatosensory cortex. We use this dataset to evaluate the quality of the cell-types we discover.

The single cell RNA-seq data, $G :=[g^1|g^2|\ldots|g^{N_C}]\in \mathbb{R}^{N_G\times N_C}$,  contains the gene expression profiles for $N_C=1691$ cells. We infer the cell types $h^i$ for these cells using equation~\eqref{eq:inferrence}, and then compute the likelihood $L^i$ of observing each for each cell under our estimated cell-type dependent gene expression profile matrix $\beta$ using equation~\eqref{eq:loglikelihood}. We can then evaluate the perplexity, a commonly used measure of goodness of fit under the vLDA model, of single cell RNA-seq data on the model we learned from our spatial point process data.

The perplexity score is a standard metric, which is defined as the geometric mean per-cell likelihood. It is a monotonically decreasing function of the log-likelihood $\mathcal{L}(G)$ of test data $G$. 
\begin{equation}
\text{perplexity}(G)= \exp(-\frac{\sum_{i=1}^{N_C} \log p({g }^i)}{\sum_{i=1}^{N_C}L^i})
\end{equation}
where the likelihood is evaluated as
\begin{equation}\label{eq:loglikelihood}
p(g^m|h^{m}, \alpha, \beta) =\frac{\Gamma\left(\sum_i\alpha_i\right)}{\prod_i \Gamma\left(\alpha_i\right)} \prod_{i=1}^{k} {(h^{m}_i)}^{\alpha_i-1} \prod\limits_{j=1}^{L^m} \left(\sum_{i=1}^{k}\sum\limits_{n=1}^{N_G} \delta_{g^{m}_j,e^n}{h^{m}_i \beta^i_n}\right).
\end{equation}
where $\delta_{i,j}$ is the Kronecker delta, $\delta_{i,j}=1$ when $i=j$ and $0$ otherwise. $e^n$ is the $n\tha$ basis vector. 

\subsection{Comparison to Standard Average Gene Expression Features Baseline and a Permutation Test for Significance}
 \begin{figure}[!htb]
\begin{center}
\begin{minipage}{0.49\textwidth}
\bc 
\psfrag{Gad1}[rC]{\scriptsize{Gad1}}
\psfrag{Sp8}[rC]{\scriptsize{Sp8}}
\psfrag{Tox3}[rC]{\scriptsize{Tox3}}
\psfrag{Nkx2-1}[rC]{\scriptsize{Nkx2-1}}
\psfrag{Lhx6}[rC]{\scriptsize{Lhx6}}
\psfrag{Pax6}[rC]{\scriptsize{Pax6}}
\psfrag{Dlx5}[rC]{\scriptsize{Dlx5}}
\psfrag{Arx}[rC]{\scriptsize{Arx}}
\psfrag{Dlx2}[rC]{\scriptsize{Dlx2}}
\psfrag{Dlx1}[rC]{\scriptsize{Dlx1}}
\psfrag{Elavl2}[rC]{\scriptsize{Elavl2}}
\psfrag{Sp9}[rC]{\scriptsize{Sp9}}
\psfrag{Tbr1}[rC]{\scriptsize{Tbr1}}
\psfrag{Foxp2}[rC]{\scriptsize{Foxp2}}
\psfrag{Tshz2}[rC]{\scriptsize{Tshz2}}
\psfrag{Stat4}[rC]{\scriptsize{Stat4}}
\psfrag{Ascl1}[rC]{\scriptsize{Ascl1}}
\psfrag{Cux2}[rC]{\scriptsize{Cux2}}
\psfrag{Neurod1}[rC]{\scriptsize{Neurod1}}
\psfrag{Mef2c}[rC]{\scriptsize{Mef2c}}
\psfrag{Oligodendrocytes}[cc]{\scriptsize{Oligodendrocytes}}
\psfrag{Interneurons}[cc]{\scriptsize{Interneurons}}
\psfrag{S1Pyramial}[cc]{\scriptsize{S1Pyramidal}}
\psfrag{Astrocytes}[cc]{\scriptsize{Astrocytes}}
\psfrag{CA1Pyramidal}[cc]{\scriptsize{CA1Pyramidal}}
\psfrag{Ependymal}[cc]{\scriptsize{Ependymal}}
\psfrag{Microglia}[cc]{\scriptsize{Microglia}}
\psfrag{Endothelial}[cc]{\scriptsize{Endothelial}}
\psfrag{Mural}[cc]{\scriptsize{Mural}}
\includegraphics[height=2.2in]{\fighome/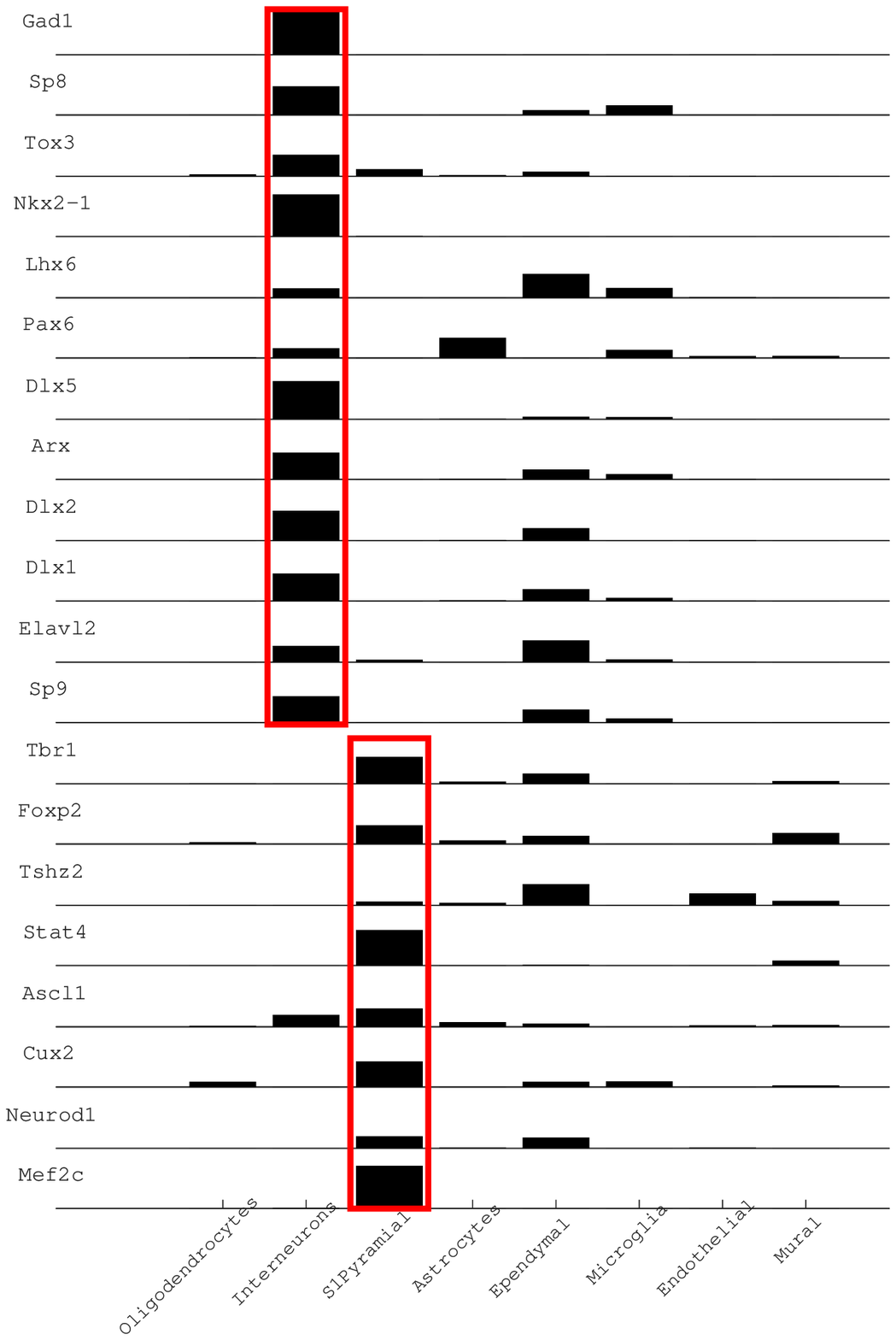}
\ec\end{minipage}
\hfil
\begin{minipage}{0.49\textwidth}
\bc
\psfrag{Spink8}[rC]{\scriptsize{Spink8}}
\psfrag{Lhx9}[rC]{\scriptsize{Lhx9}}
\psfrag{Lmo1}[rC]{\scriptsize{Lmo1}}
\psfrag{Ptrf}[rC]{\scriptsize{Ptrf}}
\psfrag{Cldn5}[rC]{\scriptsize{Cldn5}}
\psfrag{Maf}[rC]{\scriptsize{Maf}}
\psfrag{Hcls1}[rC]{\scriptsize{Hcls1}}
\psfrag{Spi1}[rC]{\scriptsize{Spi1}}
\psfrag{Myb}[rC]{\scriptsize{Myb}}
\psfrag{Fhl1}[rC]{\scriptsize{Fhl1}}
\psfrag{Aldoc}[rC]{\scriptsize{Aldoc}}
\psfrag{Sall3}[rC]{\scriptsize{Sall3}}
\psfrag{Sox21}[rC]{\scriptsize{Sox21}}
\psfrag{Mbp}[rC]{\scriptsize{Mbp}}
\psfrag{Etv6}[rC]{\scriptsize{Etv6}}
\psfrag{Sox10}[rC]{\scriptsize{Sox10}}
\psfrag{St18}[rC]{\scriptsize{St18}}
\psfrag{Olig2}[rC]{\scriptsize{Olig2}}
\psfrag{Oligodendrocytes}[cc]{\scriptsize{Oligodendrocytes}}
\psfrag{Interneurons}[cc]{\scriptsize{Interneurons}}
\psfrag{S1Pyramial}[cc]{\scriptsize{S1 Paramidal}}
\psfrag{Astrocytes}[cc]{\scriptsize{Astrocytes}}
\psfrag{CA1Pyramidal}[cc]{\scriptsize{CA1 Pyramidal}}
\psfrag{Ependymal}[cc]{\scriptsize{Ependymal}}
\psfrag{Microglia}[cc]{\scriptsize{Microglia}}
\psfrag{Endothelial}[cc]{\scriptsize{Endothelial}}
\psfrag{Mural}[cc]{\scriptsize{Mural}}
\includegraphics[height=2.2in]{\fighome/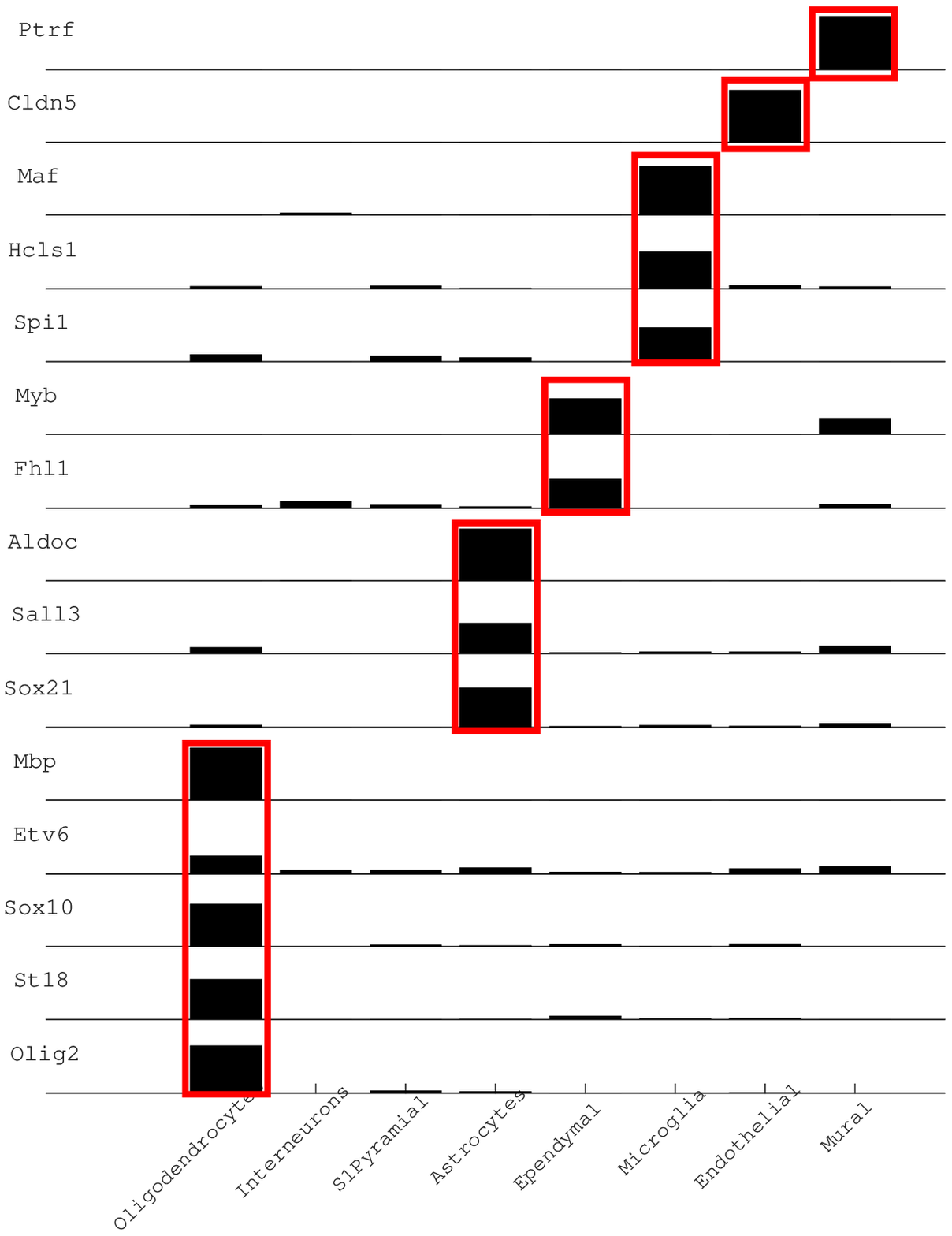}
\ec\end{minipage}\\
\end{center}
\vspace{0.5em}
\caption{Estimated memberships $\beta$ on marker genes for 8 cell types. These marker genes are used to label the columns of the membership matrix.}
%
%
\label{fig:beta}
\end{figure}

Here we demonstrate the superiority of our method and its statistical significance in two ways. First we compared the perplexity of the single-cell RNA seq dataset G under our model (figure~\ref{fig:science}, solid blue) against the perplexity of a surrogate dataset with the same marginal statistics, but whose gene-cell correlations were destroyed (figure~\ref{fig:science}, dashed blue). We generated this surrogate dataset by randomly permuting the gene expression levels for each gene across cells. This permuted dataset had a significantly higher (worse) perplexity than the true single-cell dataset. This demonstrates that our model trained to un-mix the ISH-derived spatial point processes discovered cell-types whose gene expression profiles are significantly better match to single-cells than by chance.

We also compared the predictions of cell-type gene expression profiles derived by un-mixing our spatial point process features against gene expression profiles derived by un-mixing the more standard $250\mu m \times 250\mu m \times 250\mu m$ averaged gene expression level features. We see a very large improvement in perplexity by switching from the standard simple averaging of gene expression, to extracting spatial point process features (figure~\ref{fig:science}). The single-cell RNA seq dataset analysis from figure~\ref{fig:science} shows that the perplexity of our recovered cell-types rapidly flattens after we recover approximately 10 clusters ($K=10$).

\subsection{A Brief Analysis of Recovered Cell Types in Somatosensory Cortex}

 In this section we describe the representative spatial point process statistics and gene expressions for 8 cell-types we recovered. We attempted to align our 8 clusters to cell-types defined by \cite{zeisel2015cell} in the single-cell RNA sequencing paper. We found high overlap in the gene expression profiles for all 8 clusters with known cell-types defined in~\cite{zeisel2015cell},  \emph{Interneurons}, \emph{S1 Pyramidal}, \emph{Mural}, \emph{Endothelial}, \emph{Microglia}, \emph{Ependymal}, \emph{Astrocytes} and \emph{Oligodendrocytes}, in Figure~\ref{fig:beta}.

\begin{figure}[!htb]
\begin{center}
\subfloat[a][Cell diameter in principal axes]
{\begin{minipage}{0.249\textwidth}
\bc 
\psfrag{[2,2]}[cc]{\tiny{[2,2]}}
\psfrag{[2,3]}[cc]{\tiny{[2,3]}}
\psfrag{[2,4]}[cc]{\tiny{[2,4]}}
\psfrag{[3,3]}[cc]{\tiny{[3,3]}}
\psfrag{[3,4]}[cc]{\tiny{[3,4]}}
\psfrag{[3,5]}[cc]{\tiny{[3,5]}}
\psfrag{[3,6]}[cc]{\tiny{[3,6]}}
\psfrag{[3,7]}[cc]{\tiny{[3,7]}}
\psfrag{[4,4]}[cc]{\tiny{[4,4]}}
\psfrag{[4,5]}[cc]{\tiny{[4,5]}}
\psfrag{[4,6]}[cc]{\tiny{[4,6]}}
\psfrag{[4,7]}[cc]{\tiny{[4,7]}}
\psfrag{[5,7]}[cc]{\tiny{[5,7]}}
\psfrag{[6,6]}[cc]{\tiny{[6,6]}}
\psfrag{[6,7]}[cc]{\tiny{[6,7]}}
\psfrag{[7,7]}[cc]{\tiny{[7,7]}}
\psfrag{[7,8]}[cc]{\tiny{[7,8]}}
\psfrag{[7,9]}[cc]{\tiny{[7,9]}}
\psfrag{[8,8]}[cc]{\tiny{[8,8]}}
\psfrag{[8,9]}[cc]{\tiny{[8,9]}}
\psfrag{[9,9]}[cc]{\tiny{[9,9]}}
\psfrag{Oligodendrocytes}[lB]{\tiny{Oligodendrocytes}}
\psfrag{Interneurons}[lB]{\tiny{Interneurons}}
\psfrag{S1Pyramdial}[lB]{\tiny{S1Pyramdial}}
\psfrag{Astrocytes}[lB]{\tiny{Astrocytes}}
\psfrag{CA1Pyramidal}[lB]{\tiny{CA1Pyramidal}}
\psfrag{Ependymal}[lB]{\tiny{Ependymal}}
\psfrag{Microglia}[lB]{\tiny{Microglia}}
\psfrag{Endothelial}[lB]{\tiny{Endothelial}}
\psfrag{Mural}[lB]{\tiny{Mural}}
\psfrag{Cell Fraction}[cB]{\tiny{Cell Fraction}}
\psfrag{[Diameter 1, Diamter 2]}[cc]{}
\psfrag{Diameter}[cB]{\tiny{}}
\psfrag{2}[cc]{\tiny{2}}
\psfrag{3}[cc]{\tiny{3}}
\psfrag{4}[cc]{\tiny{4}}
\psfrag{5}[cc]{\tiny{5}}
\psfrag{6}[cc]{\tiny{6}}
\psfrag{7}[cc]{\tiny{7}}
\psfrag{8}[cc]{\tiny{8}}
\psfrag{9}[cc]{\tiny{9}}
\psfrag{First principal direction}[lb]{\tiny{Axis 1}}
\psfrag{Second principal direction}[lb]{\tiny{Axis 2}}
\includegraphics[width=\textwidth]
{\fighome/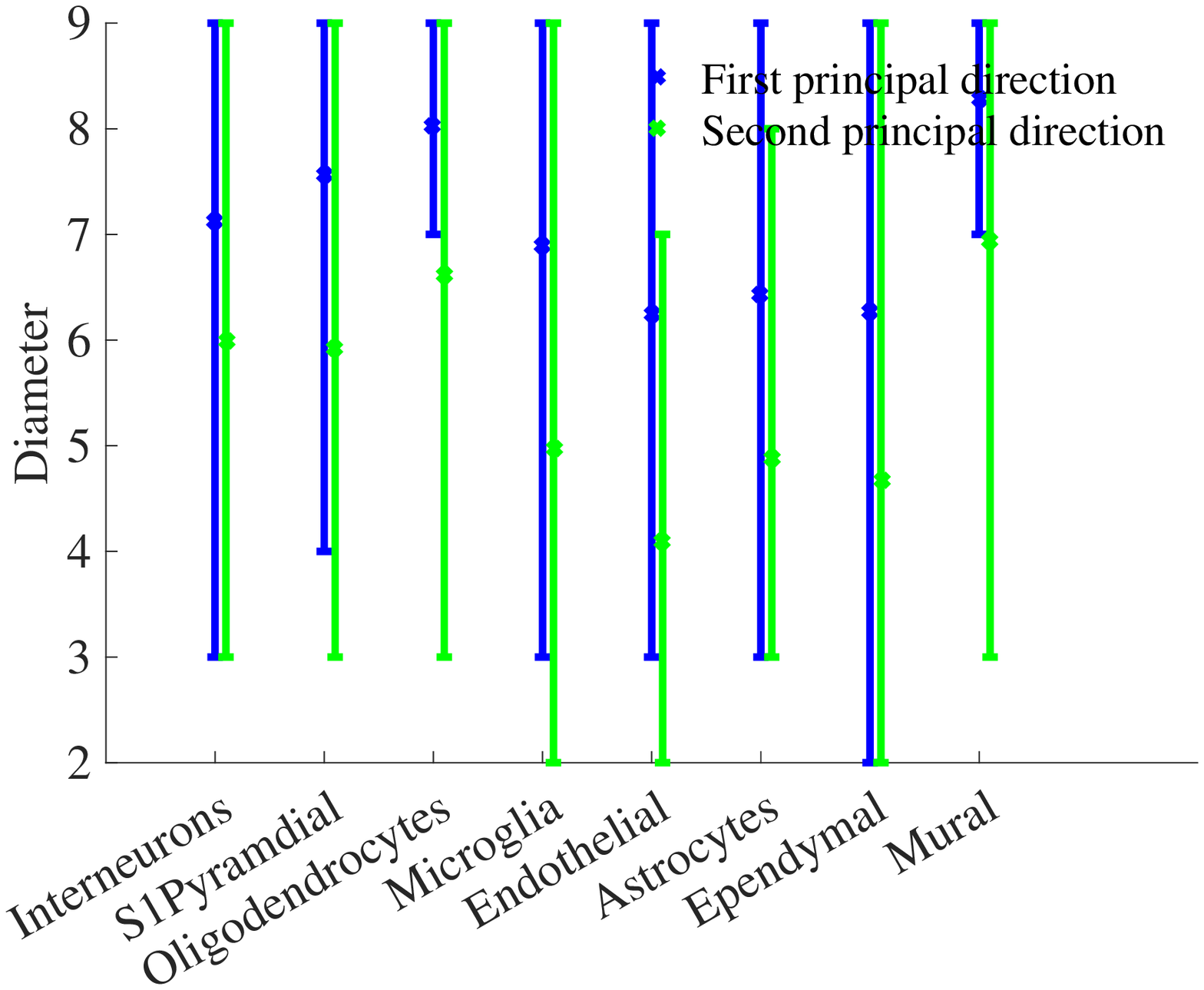}
\ec\end{minipage}}\hfil
\subfloat[b][Orientation]
{\begin{minipage}{0.249\textwidth}\bc
\psfrag{Cell Fraction}[cB]{\tiny{Cell Fraction}}
\psfrag{Orientation}[cc]{\tiny{ }}
\psfrag{0}[cl]{\tiny{0}}
\psfrag{50}[cl]{\tiny{50}}
\psfrag{100}[cl]{\tiny{100}}
\psfrag{150}[cl]{\tiny{150}}
\psfrag{pi/6}[cc]{\tiny{$\frac{\pi}{6}$}}
\psfrag{2pi/6}[cc]{\tiny{$\frac{2\pi}{6}$}}
\psfrag{3pi/6}[cc]{\tiny{$\frac{3\pi}{6}$}}
\psfrag{4pi/6}[cc]{\tiny{$\frac{4\pi}{6}$}}
\psfrag{5pi/6}[cc]{\tiny{$\frac{5\pi}{6}$}}
\psfrag{Oligodendrocytes}[lB]{\tiny{Oligodendrocytes}}
\psfrag{Interneurons}[lB]{\tiny{Interneurons}}
\psfrag{S1Pyramdial}[lB]{\tiny{S1Pyramdial}}
\psfrag{Astrocytes}[lB]{\tiny{Astrocytes}}
\psfrag{CA1Pyramidal}[lB]{\tiny{CA1Pyramidal}}
\psfrag{Ependymal}[lB]{\tiny{Ependymal}}
\psfrag{Microglia}[lB]{\tiny{Microglia}}
\psfrag{Endothelial}[lB]{\tiny{Endothelial}}
\psfrag{Mural}[lB]{\tiny{Mural}}
\includegraphics[width = \textwidth]
{\fighome/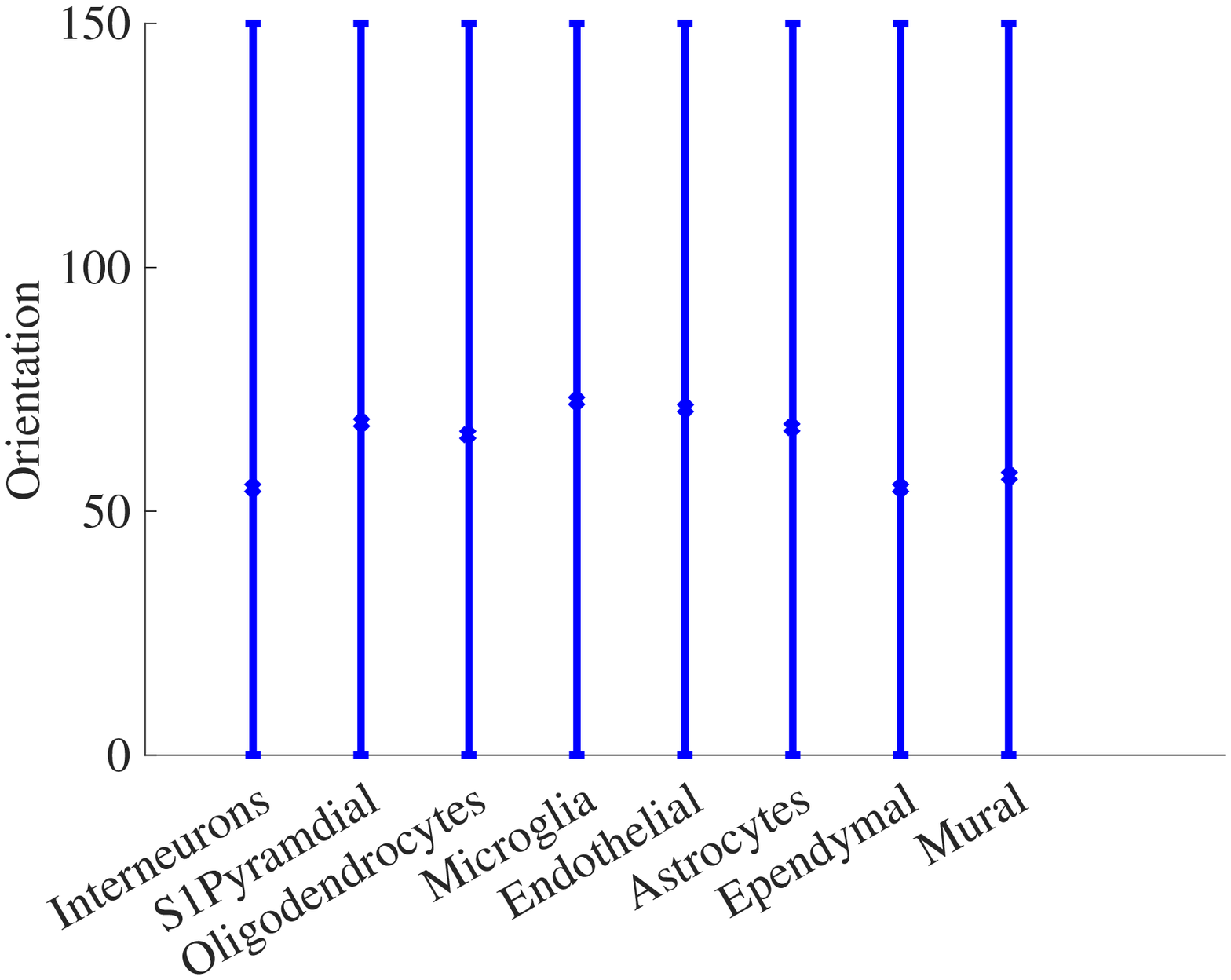}
\ec\end{minipage}}
\subfloat[c][Intensity]
{\begin{minipage}{0.249\textwidth}\bc 
\psfrag{Cell Fraction}[cB]{\tiny{Cell Fraction}}
\psfrag{Gene Intensity}[cc]{}
\psfrag{Intensity}[cB]{\tiny{}}
\psfrag{0.2}[lB]{\tiny{0.2}}
\psfrag{0.3}[lB]{\tiny{0.3}}
\psfrag{0.4}[lB]{\tiny{0.4}}
\psfrag{0.5}[lB]{\tiny{0.5}}
\psfrag{0.6}[lB]{\tiny{0.6}}
\psfrag{0.7}[lB]{\tiny{0.7}}
\psfrag{0.8}[lB]{\tiny{0.8}}
\psfrag{0.9}[lB]{\tiny{0.9}}
\psfrag{Oligodendrocytes}[lB]{\tiny{Oligodendrocytes}}
\psfrag{Interneurons}[lB]{\tiny{Interneurons}}
\psfrag{S1Pyramdial}[lB]{\tiny{S1Pyramdial}}
\psfrag{Astrocytes}[lB]{\tiny{Astrocytes}}
\psfrag{CA1Pyramidal}[lB]{\tiny{CA1Pyramidal}}
\psfrag{Ependymal}[lB]{\tiny{Ependymal}}
\psfrag{Microglia}[lB]{\tiny{Microglia}}
\psfrag{Endothelial}[lB]{\tiny{Endothelial}}
\psfrag{Mural}[lB]{\tiny{Mural}}
\includegraphics[width=\textwidth]{\fighome/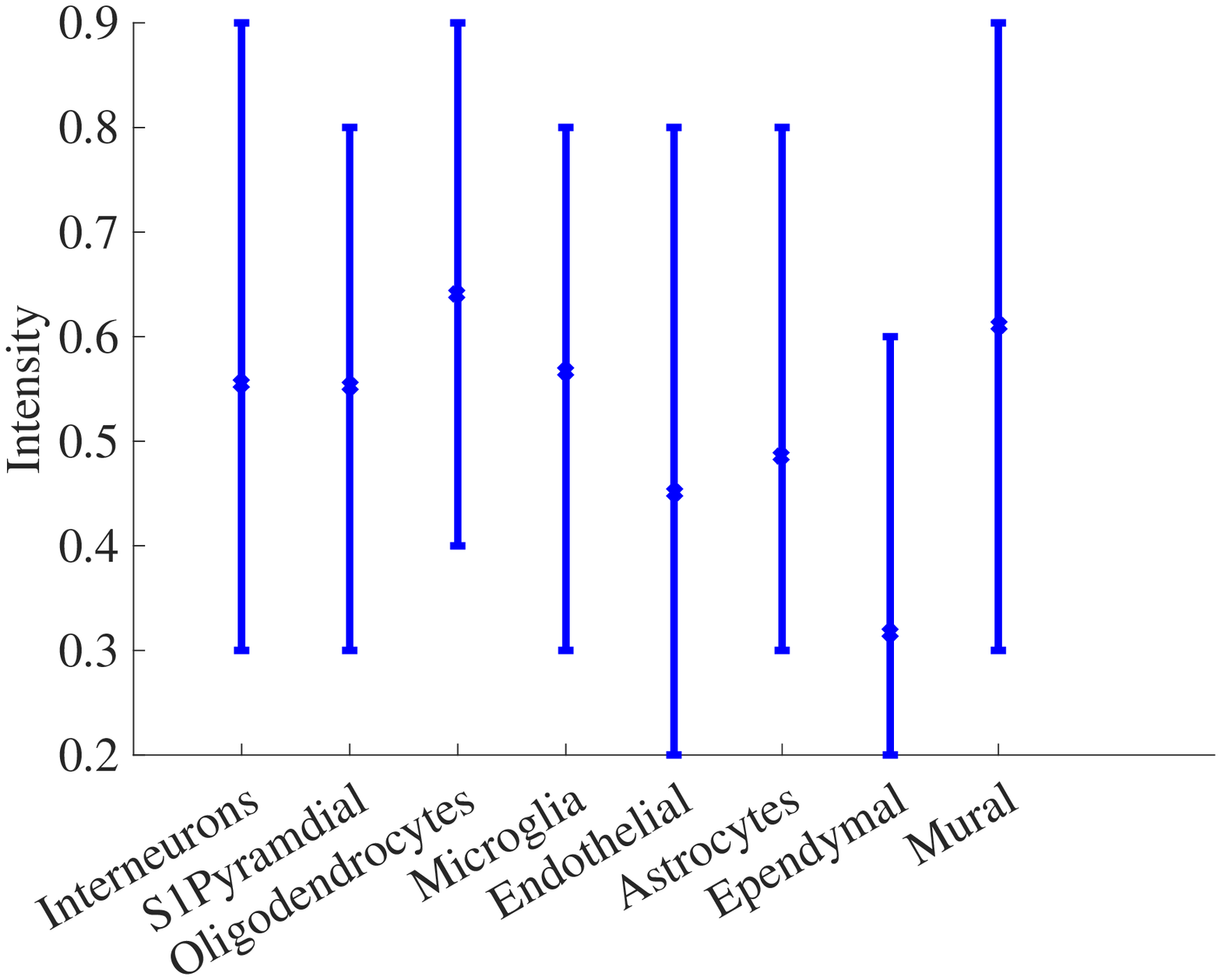}
\ec\end{minipage}}\hfil
\subfloat[d][Cells in \SI{100}{\micro\metre} radius]
{\begin{minipage}{0.249\textwidth}\bc 
\psfrag{Cell Fraction}[cB]{\tiny{Cell Fraction}}
\psfrag{Cells in 100 um}[cc]{}
\psfrag{0}[lB]{\tiny{0}}
\psfrag{10}[lB]{\tiny{10}}
\psfrag{20}[lB]{\tiny{20}}
\psfrag{30}[lB]{\tiny{30}}
\psfrag{40}[lB]{\tiny{40}}
\psfrag{50}[lB]{\tiny{50}}
\psfrag{Oligodendrocytes}[lB]{\tiny{Oligodendrocytes}}
\psfrag{Interneurons}[lB]{\tiny{Interneurons}}
\psfrag{S1Pyramdial}[lB]{\tiny{S1Pyramdial}}
\psfrag{Astrocytes}[lB]{\tiny{Astrocytes}}
\psfrag{CA1Pyramidal}[lB]{\tiny{CA1Pyramidal}}
\psfrag{Ependymal}[lB]{\tiny{Ependymal}}
\psfrag{Microglia}[lB]{\tiny{Microglia}}
\psfrag{Endothelial}[lB]{\tiny{Endothelial}}
\psfrag{Mural}[lB]{\tiny{Mural}}
\includegraphics[width=\textwidth]{\fighome/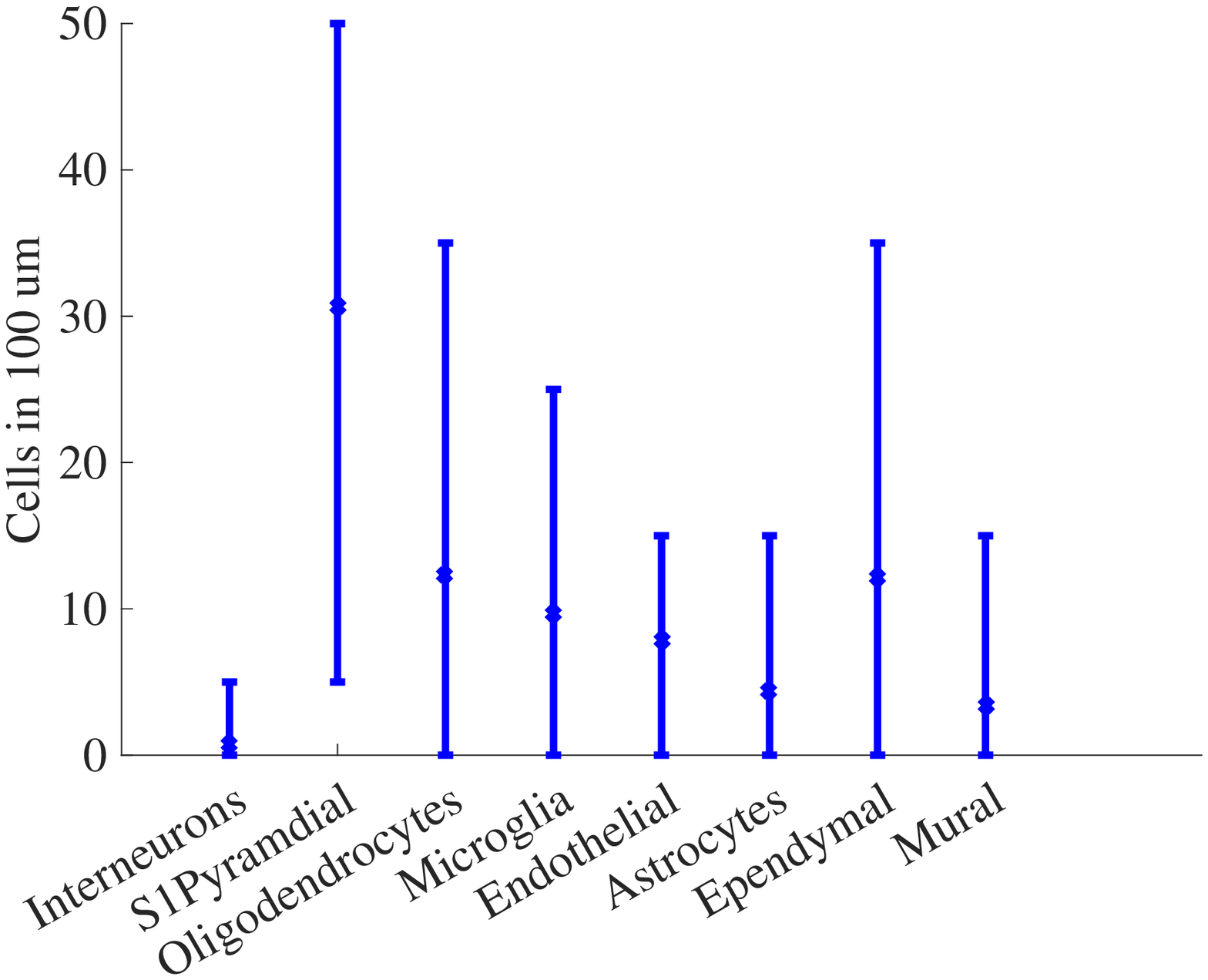}
\ec\end{minipage}}\\
\end{center}
\caption{Figure of 5\% and 95\% percentile estimated cell features for 8 cell types we detected. Inference is performed on the Spatial point process histograms data we estimated.}
\label{fig:hist_CellType}
\end{figure}

The estimate of $\beta$ was combined with MLE to infer the cell-type specific spatial point process representation $h^m_l$. 
 In examining the spatial point process distributions that we predict for each of these cell types, we discover that while the distribution of cell body orientations is quite broad and similar across cell types, the cell count distribution, which is a measure of cell density, varies in a systematic way from one cell type to another. Fig 4d shows that inhibitory Interneurons are less dense than S1Pyramidal neurons. This is consistent with their known prevalence, roughly 20\% of all neurons are GABAergic interneurons \cite{Markram:2004ek}, while the remaining 80\% are excitatory glutamatergic pyramidal neurons. As expected, this excitatory neuronal category of S1Pyramidal is the most common and hence most dense class of neuronal cells. They also have slightly larger cell bodies, compared to interneurons, as can be seen in Fig~\ref{fig:hist_CellType}a. The remaining 6 cell types correspond to various glial sub-types.
%
%

\section{Conclusion}
We developed a computational method for discovering cell types in a brain region by analyzing the high-resolution \emph{in situ} hybridization image series from the Allen Brain Atlas. Under the assumption that cell types have unique spatial distributions and gene expression profiles, we used a varied latent Dirichlet allocation (vLDA) based on spatial point process process mixture model to simultaneously infer the cell feature spatial distribution and gene expression profiles of cell types. By comparing our gene expression profile predictions to a single-cell RNA sequencing dataset, we demonstrated that our model improves significantly on state of the art.

The accuracy of our method relies heavily on the assumption that cell-types differ in their spatial distribution, and that our point process features perform a good job of distinguishing these differences. Thus the performance of our method can be improved by better estimates of better features. We would expect our method to perform better for large brain areas, which can be more accurately aligned, and which have more cells to estimate point process features.

There are several modifications to our vLDA model which might improve the faithfulness of our generative model to the biology. 
We place a symmetric Dirichlet prior over cell-type multinomial distribution $h^m$ for a given histogram bin $m$. This assumes that the number of cell-types expressing each gene is the same for all genes. But since some genes are expressed more commonly and non-specifically than others, we might expect a gene-specific prior to be a better model. Further, the symmetric Dirichlet assumes that all cell-types have equal proportions of cells. But evidence suggests that excitatory neurons are more common than inhibitory neurons in cortex~\cite{harris2013cortical}, and using a non-uniform Dirichlet prior could account for this.



{\bibliographystyle{plain}\bibliography{../nips2016_BrainMain}}
\newpage
\appendix
\begin{center}{\Large \bf Appendix for Discovering Neuronal Cell Types and Their Gene Expression Profiles Using a Spatial Point Process Mixture Model} \end{center}
\section{Morphological Basis Extraction}\label{sec:cellExtraction}
We aim to characterize the morphological basis for all cells with different size, orientation, expression profiles and spatial distribution. The traditional sparse coding introduces too many free parameters and is not suitable for compact morphological basis learning. We instead propose Gaussian prior convolutional  sparse coding (GPCSC).  The intuition for using convolution is due to the frequent replication of cells of similar shapes and the translation invariance property. 
Traditional sparse coding would learn both the shape of the cell and the location of the cell. But the convolutional sparse coding would only learn the shape here. We  characterize cell spatial distribution via decoding the sparse activation map. 

To formulate the problem formally: let $\image$ be the image observed, then the convolutional sparse coding model generates observed image $\image$ using filters (resembling cell shapes)$\filter$ superposed at locations indicated by the activation map $\map$ (whose sparsity pattern indicates cell spatial distribution and activation amplitude indicates gene expression profiles. )

Our goals of segmenting cells, extracting cell basis, and estimating gene profiles and cell locations are reduced to this optimization learning problem: 
\begin{align}\label{eq:cSparseCode}
\min\limits_{\filter_m, \map_m^{n}}   \left\| \sum\limits_{n} \image^{n} - \sum\limits_{m=1}^{k}  \filter_m \star \map_m^{n} \right\|_\fro^2 + \sum\limits_{n} \sum\limits_{m}\lambda \left\| \map_m^{n}\right\|_0,
\text{s.t.  }  &\filter_m(x,y)\ge 0,  \left\|\filter_m\right\|_\filter^2 = 1, \map_m^{(n)}(x,y)\ge 0. 
\end{align}
where $\image^{n}$ is the $n\tha$ image associated with the gene we are interested in with $D_x\times D_y$ pixels, i.e., $\image^{n}\in \mathbb{R}^{D\times D}$. 
		
We call the $F_m \in \mathbb{R}^{d\times d}$ filter, where $d$ is set to capture the local cell morphological information. The spatial coefficient for image $\image^n$ is denoted as $H_m^{(n)}\in \mathbb{R}^{(D-d+1) \times (D-d+1)}$ which represents the position of the filter $F_m $ being active on  image $\image^n$. More precisely, if $H_m^{n}(x,y) = 1$, then $F_m$ is active at $\image^{n}(x:x+d-1, y:y+d-1) $.

\subsection{Gaussian Prior Convolutional Sparse Coding}
The popular alternating approach between matching pursuit to learn activation map $\map$ and k-SVD to learn $\filter$  is general applicable to any object detection problem in image processing. However, this approach causes inexact cell number estimation as filters with multi-modality (i.e., multiple cells) are learnt.  We resolve this issue by proposing an Gaussian probability density function prior on the filters to guarantee single cell detection and achieve accurate cell number estimation. The support of $\map$ is also limited to the local maxima indicating cell centers. Note that our cell are not donut shaped, and it is reasonable to assume the darkest point being the cell center. 

Therefore, we optimize over the  objective 
$\min \left\| \sum_{n} \image^{n} - \sum_{m}  \filter_m\star \map_m^{n}\right\|_2^2+ \sum_{n} \sum_{m}\lambda \left\| \map_m^{n}\right\|_0$ such that $\filter_m$ are $2-D$ Gaussian densities with priori set top 2 principal radius and orientation. 
Alternating Minimization is used to solving the optimization problem. If we define the residual as $\sum_n\image^n -  \sum_n\sum_{m} \widehat{\filter}_m\star \widehat{\map}_m^{n}$, the gradient of the objective reduced to an iterative approach of updating filters, compute residual, optimizing activation map based on residual, compute residual and updating filters again. 
	It is easy to see that both $\frac{\partial L}{\partial F_m}(i,j)$ and $\frac{\partial L}{\partial H_m}(i,j)$ are convolution of the residual and the other variable rotated by angle $\pi$.

\subsection{Image Registration/Alignment}
A structure represents a neuronanatomical region of interest. Structures are grouped into ontologies and organized in a hierarchy or structure graph. 
We are interested in the somatosensory cortex area. So we use the affine transform from Allen Brain Institute~\cite{AMBA, lein2007genome} to align all the in-situ hybridization images with the Atlas brain to extract the correct region. 

\end{document}